\documentclass[12pt]{article}
\usepackage{amsmath,amsfonts,amssymb}
\usepackage{fancyhdr,graphicx,amsmath, amscd}
\usepackage{color}
\def\be{\begin{equation}}
\def\ee{\end{equation}}
\def\bea{\begin{eqnarray}}
\def\eea{\end{eqnarray}}
\def\bt{\begin{theorem}}
\def\et{\end{theorem}}
\def\bl{\begin{lemma}}
\def\el{\end{lemma}}
\def\br{\begin{remark}}
\def\er{\end{remark}}
\def\bc{\begin{corollary}}
\def\ec{\end{corollary}}
\def\bd{\begin{definition}}
\def\ed{\end{definition}}
\def\a{\alpha}
\def\g{\gamma}
\def\d{\delta}

\def\l{\lambda}

\def\r{\rho}
\def\s{\sigma}
\def\t{\tau}

\def\G{\Gamma}
\def\D{\Delta}

\def\L{\Lambda}

\def\cD{\mathcal{D}}
\def\cE{\mathcal{E}}
\def\cF{\mathcal{F}}

\def\cH{\mathcal{H}}

\def\cK{\mathcal{K}}
\def\cL{\mathcal{L}}

\def\cS{\mathcal{S}}

\def\cU{\mathcal{U}}

\def\bbC{\mathbb{C}}

\def\bbP{\mathbb{P}}

\def\bbR{\mathbb{R}}

\def\b1{B_{1}^}

\def\ba{\begin{array}}
\def\ea{\end{array}}
\def\ben{\begin{enumerate}}
\def\een{\end{enumerate}}
\newtheorem{theorem}{Theorem}
\newtheorem{lemma}{Lemma}
\newtheorem{remark}{Remark}
\newtheorem{corollary}{Corollary}

\newtheorem{definition}{Definition}
\begin{document}
\title{On the comparison of volumes of quantum states
}
\author{Deping Ye\thanks{Department of Mathematics, 202 Mathematical Sciences
Bldg, University of Missouri Columbia, MO 65211 USA. {\tt
email:deping.ye@gmail.com}.}}
\date{ }
\maketitle
\begin{abstract}
This paper aims to study the $\a$-volume of $\cK$, an arbitrary
subset of the set of $N\times N$ density matrices. The $\a$-volume
is a generalization of the Hilbert-Schmidt volume and the volume
induced by partial trace. We obtain two-side estimates for the
$\a$-volume of $\cK$ in terms of its Hilbert-Schmidt volume. The
analogous estimates between the Bures volume and the $\a$-volume
are also established. We employ our results to obtain bounds for
the $\a$-volume of the sets of separable quantum states and of
states with positive partial transpose (PPT). Hence, our
asymptotic results provide answers for questions listed on page 9
in \cite{K. Zyczkowski1998} for large $N$ in the sense of
$\a$-volume.

\end{abstract}

\section{Introduction}
Recent development of quantum information and quantum computation
theory has attracted considerable attention on the geometry of
quantum states, which studies the geometry of the set and/or
subsets of quantum states on the (composite) Hilbert space $\cH$
(of complex dimension $N$). Among those subsets, the set of all
quantum states ($\cD$), the set of separable quantum states
($\cS$), the set of entangled quantum states ($\cE:=\cD\setminus
\cS$), and the set of quantum states with positive partial
transpose ($\mathcal{PPT}$) are of particular interest. In
applications (see \cite{RevModPhys.81.865, Nie1}), entangled
quantum states play fundamental roles because of the
Einstein-Podolsky-Rosen (EPR) correlations \cite{EPR1} (see also
\cite{Sch1}). Hence it is important to study (1) the probability
of finding separable quantum states within $\cD$; (2) the
necessary and/or sufficient conditions of a quantum state being
separable (or entangled). Unfortunately, it turns out that the
second problem is hard \cite{Gur}. One important and powerful {\em
necessary condition} for separability is the well-known
Peres-Horodecki positive partial transpose (PPT) criterion
\cite{Pe1}. However, it is known that the Peres-Horodecki PPT
criterion is not sufficient in general (except in very special
cases; for instance, \cite{Ho1, ST1, Wor1}), and entangled quantum
states with PPT have been constructed \cite{Ho2}. Thus, a natural
question is (3) {\it how precise is the Peres-Horodecki PPT
criterion as tools to detect the separability?}  Another question
regarding to the Peres-Horodecki PPT criterion is (4) {\it is the
Peres-Horodecki PPT criterion precise as a tool to detect
entanglement?} To answer questions (1), (3) and (4), one often
needs to study the size of $\cE$, $\cS$, $\mathcal{PPT}$, and
$\cD$ (for various relevant measures of size).

\vskip 2mm In literature, measures on $\cD$ have one common
feature: unitary invariance. These measures can be written as the
product of measures on simplex (of eigenvalues) and the measure on
the manifold (of eigenvectors). Important measures on $\cD$
include, for instance, the Hilbert-Schmidt measure ($V_{HS}$), the
Bures measure ($V_B$), and the measure induced by partial trace on
composite systems. The Hilbert-Schmidt measure is induced by the
Hilbert-Schmidt metric which induces the flat and Euclidean
geometry into $\cD$. Hence, a lot of known techniques, such as,
techniques from geometric functional analysis and convex geometry,
can be used to estimate the Hilbert-Schmidt volume of (convex)
subsets of $\cD$, e.g., $\cS$ and $\mathcal{PPT}$ in \cite{AS1,
Sz1}. These two papers provided answers to questions (1) and (3)
for large $N$ in the sense of Hilbert-Schmidt volume. In
\cite{YeDeping2009}, the author obtained similar results in the
sense of Bures volume, by comparing the Bures volume with the
Hilbert-Schmidt volume.

\vskip 3mm The present paper strives to answer questions (1) and
(3) for large $N$ in the sense of the $\a$-volume (see section 2
for its definition). The $\a$-volume is a natural generalization
of the Hilbert-Schmidt volume and the volume induced by partial
trace on composite systems. The latter one has appeared in many
places and attracted a lot of attention \cite{Braunsein1996,
Hall1, Lloyd1988, Lubkin1978, Page1993, Sommers2004, Zycz2}.  It
is inspired by the important procedure of purification and the (up
to a multiplicative constant) unique measure on the space of pure
states of the (complex) Hilbert space $\cH$. Any quantum state
$\r$ on $\cH$ may be obtained by partial tracing of some pure
quantum state $|\phi\rangle \langle \phi|$ over the auxiliary
subsystem $\cH'$ (with dimension $K \geq N$), where $|\phi\rangle
$ is a unit (column) vector in the composite Hilbert space $\cH
\otimes \cH'$. The space of pure states of $\cH$ is isomorphic
with the complex projective space $\bbC P^{N-1}$, and hence the
unique measure on it, $P_N(\cdot)$, is the one induced by the Haar
measure on the unitary group $\cU(N)$. Therefore, to obtain
$V_{N,K}$, the volume induced by partial trace on the composite
Hilbert space $\cH\otimes \cH'$,  one chooses the natural measure
$P_{NK} (\cdot)$ on the space of pure states of $\cH\otimes \cH'$,
and considers its push-forward induced by the the operation of
partial trace \cite{Zycz2}. When $K=N$, $V_{N,K}$ coincides with
the Hilbert-Schmidt measure. Thanks to the work of Sommers and
\.Zyczkowski \cite{Sommers2004, Zycz2}, one knows the precise
mathematical formula of $V_{N,K}$ and some statistical properties
of $V_{N,K}$. In particular, they calculated the exact value of
$V_{N,K}(\cD)$ for all $N$ and $K$. However, the calculation of
the exact values of $V_{N,K}(\cS)$ and $V_{N,K}(\mathcal{PPT})$
seems to be difficult because the geometry of these sets is not
very well understood. These quantities can be used to measure the
probabilities of separability and of PPT within $\cD$. In the
present paper, we provide estimates of $V_{N,K}(\cS)$ and
$V_{N,K}(\mathcal{PPT})$, as well as the $\a $-volumes of $\cS$
and $ \mathcal{PPT}.$ {\it Our results show that the probability
of finding separable states in $\cD$ is extremely small and so the
Peres-Horodecki PPT criterion is not precise for (even moderate)
large $N$, in the sense of $\a$-volume.}

\vskip 3mm The present paper is organized as follows. In section
2, we introduce some necessary mathematical background and
notations; all the results cited there are known or
straightforward. Our main results are presented in section 3, that
is, we compare the $\a$-volume with the Hilbert-Schmidt volume and
the Bures volume. Section 4 contains estimates of the $\a$-volume
of $\cS$ and $\mathcal{PPT}$, and these new estimates supply
solutions for questions (1) and (3) for large $N$. Moreover, we
provide some numerical examples to show the effectiveness of our
new estimates.

\section{Mathematical Background and Notations}

\vskip 2mm   We work on the (complex) Hilbert space $\cH=\bbC
^{D_1}\otimes \bbC ^{D_2}\cdots \otimes \bbC ^{D_n}$ with $n\geq
2$ and  $D_i\geq 2$ for all $i=1, 2, \cdots, n$. The (complex)
dimension of $\cH$ equals $N=D_1D_2 \cdots D_n$. Any quantum state
on $\cH$ can be represented as a density matrix, i.e., a $N\times
N$ positive (semi) definite matrix with trace $1$. Here the trace
of a matrix is the sum of its diagonal elements. We use
$\cD=\cD(\cH)$ to denote the set of all quantum states on $\cH$.
The set of separable quantum states on $\cH$ \cite{Wer1} is
denoted by $\cS$, that is,
$$\cS =\cS(\cH):={\rm conv}\{\rho _1 \otimes \cdots \otimes \rho _n, \rho _i\in
\cD (\bbC ^{D_i})\}.$$ The complement of $\cS $ within $\cD$ is
the set of entangled quantum states, i.e., $\cE:=\cD\setminus
\cS$. Note that the entangled quantum states play crucial roles in
quantum information and quantum computations. Both $\cD$ and $\cS$
are convex subsets with (real) dimension $d=N^2-1$.

\vskip 3mm  For any quantum state $\r\in \cD$, there are some
unitary matrix $U\in \cU(N)$ and some diagonal matrix $\L={\rm
diag}(\l _1, \cdots, \l_N)$ with $(\l_1, \cdots, \l_N)\in \D$,
such that, $\r =U\L U^\dag$. Hereafter, $A^\dag$ is the complex
conjugate of the matrix $A$, $\cU(N)$ denotes the $N$-th unitary
group, that is, the set of all $N\times N$ unitary matrices. $\D$
refers to the regular simplex in $\bbR ^N$, i.e.,
$\D=\left\{(\l_1, \cdots, \l_N)\in \bbR^N: \l_i\geq 0, \sum
_{i=1}^N \l _i=1\right\}.$ Obviously, the eigenvalue decomposition
$\r =U\L U^\dag$ is {\em not} unique. One can change the order in
which the eigenvalues of $\r$ occur by using unitary permutation
matrices. Without loss of generality, one can choose $(\l_1,
\cdots, \l_N) \in \D_1,$ the subset of $\D$ with the order
$\l_1\geq \cdots\geq \l_N\geq 0$. This corresponds to divide
$\Delta$ into $N!$ parts, and to pick just one of them. The reason
for establishing an order in the eigenvalues of $\r$ is to avoid
duplication in calculating volume of subsets of $\cD$. In the case
of non-degenerate spectrum, i.e., $\l_1>\l_2>\cdots
>\l_N$, it is easy to verify that
$$\r=U\L U^{\dag} =U B \L B^{\dag} U^{\dag},$$ where $\L=diag(\l_1, \cdots,
\l_N)$ and $B=diag(z_1, \cdots, z_N)$ with $|z_j|=1$ for $j=1,
\cdots, N$. That is, the matrix $U$ is determined up to the $N$
arbitrary phases entering $B$, and the orbit will be the coset
space $\cF ^N={\cU(N)}/{[\cU(1)]^N}$. Again to avoid duplication
in calculating volume of subsets of $\cD$, we will use measure on
$\cF ^N$ instead of measure on $\cU (N)$. Note that if
degeneracies occur in the spectrum of $\r$, the matrix $B$ need
{\em not} to be diagonal in order to commute with $\L$, however,
the degenerate spectrum can be ignored because its total measure
is $0$.

\vskip 2mm \par Measures on $\cD$, such as, the Hilbert-Schmidt
measure, the Bures measure, and the measure induced by partial
trace, are invariant under conjugation by a unitary matrix.
Moreover, they all have the product form: $\,d\nu \times \,d\g$,
where $\nu$ are some measures on the simplex $\D$ and $\g$ is the
invariant measure on $\cF ^N$. The $\g$ measure may be written as
$$d\g = \prod _{1\leq i<j\leq N} 2Re(U^{-1}{d}U)_{ij}Im(U^{-1}{d}U)_{ij},$$ where $dU$ is the variation of $U$ such that $U, U+dU\in
\cU(N)$. Note that the $\g$ measure is the unique measure (up to a
multiplicative constant) induced by Haar measure on the unitary
group $\cU(N)$. It is known that $\g(\cF^N)$, the total $\g$
measure of $\cF ^N$ (see \cite{Zycz1}), is equal to
\begin{equation} \label{gam-measure}Z_N=\displaystyle \frac{(2\pi
)^{N(N-1)/2}}{E(N)} , \quad \mbox{where $E(N)=\prod _{j=1}^N
\Gamma(j)$}.\end{equation}  Here $\G(x)=\int
_0^{\infty}t^{x-1}e^{-t}\,dt$  is the Gamma function. We point out
that \begin{equation}\label{YD3} \frac{1}{\vartheta x}\leq
\G(x)\leq \frac{1}{x},\ \ \mbox{or} \  \ \frac{1}{\vartheta }\leq
\G(1+x)\leq 1, \ \ \ \mbox{for all $x\in (0,1)$},\end{equation}
where $\vartheta\thickapprox 1.12917$ \cite{Deming1935}.

\vskip 3mm For all $\a>0$, we define the $\a$-volume, $V_{\a}$, as
\begin{eqnarray}\label{avolume:element-1}
\,d V_{\a}=\prod _{i=1}^N\l_i^{\a-1} \prod_{1\leq i<j\leq
N}(\l_i-\l_j)^2\ \,d\L\,d\g.
\end{eqnarray} Here, for simplicity, we let  $$\,d\L=\d_{0}(\sum_{i=1}^N
\l_i-1) \prod_{i=1}^{N}\,d\lambda_i,$$ where $\d_0$ is the dirac
measure at $0$. The $\a$-volume of $\cD$ can be calculated as
follows (see \cite{Zycz2, Zycz1, Ben1, MML1})
\begin{eqnarray}\!V_{\a}(\cD)\!\!=\!\!\int _{\cD}\! \prod_{1\leq i<j\leq
N}\! |\lambda_i-\lambda_j|^{2}\! \prod _{i=1}^N \lambda _i
^{\a-1}\!\,d\L \,d\g \!= \!\frac{(2\pi)^{N(N-1)/2}}{\Gamma (\a
N+N(N-1))}\! \prod_{j=1}^N \Gamma(j+\a-1). \label{avolume:D}
\end{eqnarray}

\vskip 2mm The Hilbert-Schmidt distance between any two states
$\r, \s \in \cD$ is defined as $$D_{HS}(\r,
\s)=\|\r-\s\|_{HS}=\sqrt{tr((\r-\s)^2)}.$$ This (natural) metric
equips $\cD$ with the flat, Euclidean geometry on $\cD$, but it is
not monotone \cite{Ozawa2000}. It induces the Hilbert-Schmidt
measure $V_{HS}$, which equals to $\sqrt{N}V_1$
\cite{Zycz2,Zycz1}, namely,
$$ \,dV_{HS}=\sqrt{N}\prod _{1\leq i<j\leq N}(\lambda _i-\lambda
_j)^2 \,d\Lambda  \, d\g.$$ Note that the Hilbert-Schmidt measure
is same as the usual (translation invariant) Lebesgue measure on
$\cD$. Moreover, the Hilbert-Schmidt measure {\em does not} have
singularities. By formula (\ref{avolume:D}), the precise value of
the Hilbert-Schmidt volume of $\cD$ \cite{Zycz1} equals
\begin{eqnarray} V_{HS}(\mathcal{D})&=&{(2\pi )^{\frac{N(N-1)}{2}} }\ \sqrt{N} \ \frac{E(N)}{\Gamma
(N^2)}\label{Hilbert-Schmidt-D}.\end{eqnarray} In later sections,
we are interested in the ratio $\frac{V_{HS}(\cK)}{V_{HS}(\cD)}$
for $\cK\subset \cD$, which is equal to the ratio
$\frac{V_{1}(\cK)}{V_{1}(\cD)}$. Hence, it is convenient to ignore
the constant $\sqrt{N}$ for our analysis, and we keep using
$V_{HS}$ instead of $V_1$ to represent the $1$-measure, i.e.,
\begin{eqnarray}\label{Hilbert} \,dV_{HS}=\prod _{1\leq i<j\leq
N}(\lambda _i-\lambda _j)^2 \,d\Lambda  \, d\g.\end{eqnarray}

\vskip 2mm Another special case of the $\a$-volume is the volume
induced by partial tracing of the composite quantum system
$\cH\otimes \cH'$. Here $\cH$ and $\cH'$ are $N$ and $K$
dimensional Hilbert spaces respectively. Without loss of
generality, we assume $K\geq N$. The partial tracing over $\cH'$
gives a reduced density matrix of size $N\times N$. Any state
$\rho $ on $\cH \otimes \cH'$ can be expressed uniquely as
$$\rho = \sum _{i,j}^{N} \sum _{\a, \beta }^{K} \rho _{i\a, j\beta } |e_i\otimes
f_{\a}\rangle \langle e_j\otimes f_{\beta}|,$$ where
$\{e_i\}_{i=1}^{N}$ and $\{f_\a\}_{\a=1}^{K}$ are the canonical
bases of $\cH$ and $\cH'$ respectively. Define the partial trace
over $\cH'$ as
$$\r^A=Tr_B(\r), \ \mathrm{where}\  \r_{ij}^A=\sum _{\beta=1}^{K}\r _{i\beta,
j\beta} \ \mathrm{for}\ i,j=1, \cdots, N.$$

 The measure induced
by partial trace is an alternative way to derive measures on
$\cD$. In fact, the partial trace process allows us to view states
on $\cH$ as a (pure) state on (much higher) dimensional space
$\cH\otimes \cH'$. Then the measures induced by partial trace may
be considered as a projection of the $(NK-1)$ dimensional simplex
of eigenvalues into simplex of $(N-1)$ dimension $\D$
\cite{Zycz2}. It takes the form
\begin{eqnarray*} \,d V_{N,K}=\prod _{i=1}^{N}\l_i^{K-N}
\prod_{1\leq i<j\leq N}(\l_i-\l_j)^2\ \,d\L\,d\g,
\end{eqnarray*} where $(\lambda_1, \cdots, \l _N)\in \D_1$. This is the $\a$-volume with $\a=K-N+1$, a natural number
bigger than or equal to 1, and thus, (integer) $\a$-measure is an
induced measure on $\cD$. In particular, $V_{N,N}$ is just the
Hilbert-Schmidt measure. In other words, the Hilbert-Schmidt
measure can be viewed as an induced measure on $\cD$.

\vskip 3mm  The Bures measure $V_B(\cdot)$ \cite{Hall1, Ben1,
Soz1} can be formulated as follows:
\begin{eqnarray*}
dV_B= \frac{2^{\frac{2-N-N^2}{2}}}{\sqrt{\l _1\cdots \l_N}} \prod
_{1\leq i<j\leq N}\frac{(\lambda_i-\lambda_j)^2}{\lambda_i
+\lambda_j}\,d\Lambda \,d\g,\end{eqnarray*} where $(\lambda_1,
\cdots, \l _N)\in \D_1$. It is induced by the Bures metric
$D_B(\cdot, \cdot)$ \cite{Bur1, Uh0}, which takes the following
form: for any two states $\r, \s\in \cD$, $D_{B}(\r,
\s)=\sqrt{2-2tr\sqrt{\sqrt{\r}\s\sqrt{\r}}}.$ The Bures metric is
proven to be equivalent to \cite{Barnum1996}
$$D_B(\r, \s)=\sup _{\{\bbP _i\} } \bigg(\sum _{i=1}^N
\big[\sqrt{tr(\bbP _i \r)}-\sqrt{tr(\bbP _i
\s)}\big]^2\bigg)^{\frac{1}{2}},$$ where the supremum runs over
all the projection measurements $\{\bbP_i\}_{i=1}^N$, i.e., all
$\bbP_i$ are rank $1$ projections satisfying $\sum _{i=1}^N
\bbP_i=Id_N$. Employing the projection measurement $\{\bbP
_i\}_{i=1}^N$ to a state $\r$ results in the $i$-th outcome with
probability $tr(\bbP_i\r), \ i=1, \cdots, N$. Note that $
\big(\sum _{i=1}^N \big[\sqrt{tr(\bbP _i \r)}-\sqrt{tr(\bbP _i
\s)}\big]^2\big)^{\frac{1}{2}}$ is the Hellinger distance, which
measures the statistical distinguishability between the discrete
probability distributions $\{tr(\bbP_i\r)\}_{i=1}^N$ and
$\{tr(\bbP_i\s)\}_{i=1}^N$. Hence, the Bures distance provides the
measurement of statistical distinguishability between $\r$ and
$\s$. We point out that the Bures metric is Riemannian but not
flat. Moreover, it is monotone, that is, it does not increase
under the quantum channels (completely positive, trace preserving
maps) \cite{Petz1996}. The precise value of the Bures volume of
$\cD$ was obtained in \cite{Soz1},
\begin{eqnarray}\!V_B(\cD)\!&=&\!\!\int _{\D_1\;\times \; \cF
^N}\!\! \frac{2^{\frac{2-N-N^2}{2}}}{\sqrt{\l _1\cdots \l_N}}\!\!
\prod _{1\leq i<j\leq
N}\!\!\frac{(\lambda_i-\lambda_j)^2}{\lambda_i +\lambda_j}
\,d\Lambda \,d\g \nonumber =2^{1-N^2}\frac{\pi
^{{N^2}/{2}}}{\Gamma({N^2}/{2})},\label{Bures:precise}\end{eqnarray}
which is the $d$-dimensional volume of the $d$-dimensional {\em
hemisphere} with radius $\frac{1}{2}$. (There is no satisfactory
explanation for this mysterious fact.) Noticed that the Bures
volume {\em is not} translation invariant and has singularities on
the boundary of $\cD$, where {\em at least one} of the $\l_i$'s
are $0$. ( If {\em two or more} of them are $0$, then some
denominators in (\ref{BV-1}) vanish.) As in the case of
Hilbert-Schmidt volume, it is convenient to ignore the constant
term ${2^{\frac{2-N-N^2}{2}}}$ in $\,dV_B$. Hereafter, we keep
using $\,dV_B$ to represent the measure
\begin{eqnarray}\label{BV-1} dV_B= \frac{1}{\sqrt{\l _1\cdots
\l_N}} \prod _{1\leq i<j\leq
N}\frac{(\lambda_i-\lambda_j)^2}{\lambda_i +\lambda_j}\,d\Lambda
\,d\g,\end{eqnarray}

\vskip 2mm The primary goal in \cite{YeDeping2009} is to compare
$\mathrm{VR}_{B}(\cK, \cD)$ with $\mathrm{VR}_{HS}(\cK,\cD)$ for
any subset $\cK$ in $\cD$. Here the Bures volume radii ratio,
$\mathrm{VR}_{B}(\cK, \cL)$, for two subsets $\cK, \cL\subset
\cD$, was defined as \cite{YeDeping2009}
$$\mbox{VR}_{B}(\cK,
\cL)=\left(\frac{V_{B}(\cK)}{V_{B}(\cL)}\right)^{1/d},$$ where
$d=N^2-1$ and the Hilbert-Schmidt volume radii ratio was defined
as
$$\mbox{VR}_{HS}(\cK,
\cL)=\left(\frac{V_{HS}(\cK)}{V_{HS}(\cL)}\right)^{1/d}.$$ These
quantities aim to compare the volume ratio of $\cK$ to $\cL$.
Similarly, we define here the $\a$-volume radii ratio,
$\mathrm{VR}_{\a}(\cK,\cL)$, as
$$\mbox{VR}_{\a}(\cK,
\cL)=\left(\frac{V_{\a}(\cK)}{V_{\a}(\cL)}\right)^{1/d}.$$ This
can be used as a measure of the relative size of $\cK$ to $\cL$ in
the $\a$-volume sense, and does have geometric meanings. This
paper strives to estimate $\mathrm{VR}_{\a}(\cK, \cD)$ from both
upper and below in terms of $\mathrm{VR}_{HS}(\cK,\cD)$. Our
proofs rely on the Stirling approximation, which can be written as
\begin{equation}\label{Stir:1}\Gamma(z)=\sqrt{\frac{2
\pi}{z}}~{\left(\frac{z}{e}\right)}^z
\left(1+O\left(\frac{1}{z}\right)\right).\end{equation}

\vskip 2mm \par We refer the readers to the references
\cite{YeDeping2009, Hall1, Zycz2, Zycz1, Ben1, Soz1, Hip1} for
more detailed background and for motivation.

\vskip 3mm \section{$\a$-volume VS Hilbert-Schmidt and Bures
volume} \vskip 4mm

This section aims to compare the $\a$-volume of subsets of quantum
states in terms of their Hilbert-Schmidt volume and Bures volume.
Hereafter, we let $\cK$ be any (Borel) measurable subset of
$\cD(\cH)$, the set of quantum states on (complex) Hilbert space
$\cH$. The following lemma, which is independent of possible
tensor product structures of $\cH$, is our main tool to estimate
the $\a$-volume in terms of the Hilbert-Schmidt volume. The same
approach will give corresponding comparison results for real
Hilbert space $\cH$.


\bl \label{avolume:Hilbert:lemma:1} Let $\a>0$ be a (fixed)
constant independent of the dimension of $\cH$. Let $\cH$ be any
(complex) Hilbert space with (complex) dimension $N$ and $\cK$ be
any measurable subset in $\cD(\cH)$.\vskip 3mm \noindent  (i): For
all $p>\a>1$,\begin{eqnarray}\label{avolume:Hilbert:ageq1}
{V_{HS}(\mathcal{K})}^{p}\ \bigg(V_{\big(\frac{\a-p}{1-p}\big)}
(\cD) \bigg)^{1-p} \leq V_{\a}(\mathcal{K}) \leq N^{(1-\a)N} 
\ V_{HS}(\mathcal{K}),
\end{eqnarray}

\noindent (ii): For all
$0<p<\a<1$,\begin{eqnarray}\label{avolume:Hilbert:aleq1}
N^{(1-\a)N}
\ V_{HS}(\mathcal{K}) \leq V_{\a}(\mathcal{K})\leq
{V_{HS}(\mathcal{K})}^{p}\ \bigg(V_{\big(\frac{\a-p}{1-p}\big)}
(\cD) \bigg)^{1-p}.
\end{eqnarray}
\el

\vskip 2mm \noindent {\bf Remark.} From formula (\ref{avolume:D}),
for $\frac{\a-p}{1-p}>0$, one has
\begin{eqnarray*}V_{
\frac{\a-p}{1-p}}(\cD)=\frac{(2\pi)^{N(N-1)/2}}{\Gamma
\left(\big(\frac{\a-p}{1-p}\big)N+N(N-1)\right)} \prod_{j=1}^N
\Gamma\left(j+\bigg(\frac{\a-1}{1-p}\bigg)\right).
\end{eqnarray*} The above formula, as a function of $p$, can be extended to $\frac{\a-p}{1-p}\leq 0$; however there are singularities
whose exact locations depend on $N$ and $\a$.

\vskip 3mm \par \noindent {\bf Proof.} We only prove the case of
$\a>1$. Similar approach will lead to the argument for the case of
$\a<1$.

\vskip 2mm \par To that end, let $h: \D \rightarrow \bbR$ be $h(\l
_1, \cdots, \l_N)=\prod_{i=1}^N \l _i^{\a-1}.$ Note that $\partial
\D$, the boundary of the simplex $\D$, consists of sequences for
which one or more of the $\lambda_i$'s equal to $0$. Therefore,
$h(\l _1, \cdots, \l _N)$ is always $0$ on $\partial \D$ if
$\a>1$, and is strictly positive in the interior of the simplex
$\D$. On the other hand, $h(\l _1, \cdots, \l _N)$ has unique
critical point $(1/N, \cdots, 1/N)$ in the interior of simplex
$\D$ by the Lagrange multiplier method.
 The compactness of $h(\l _1, \cdots,
\l _N)$ implies that $h(\l_1, \cdots, \l_N)$ must have a maximum
inside the interior of the simplex $\D$ for $\a>1$, and hence the
(unique) critical point $(1/N, \cdots, 1/N)$ must be the (only)
maximizer of $h(\l_1, \cdots, \l_N)$ on $\D$. Therefore,
\begin{equation}\label{Estimation-h:geq1}h(\l_1, \cdots, \l_N)=\prod _{i=1}^N \l_i^{\a-1}\leq N^{(1-\a)N}, \ \ for \ \ \a>1.\end{equation}
By formula (\ref{avolume:element-1}), the $\a$-volume of $\cK$ can
be calculated as follows:
\begin{eqnarray}\label{avolume:K}
V_{\a}(\mathcal{K})=\int _{\mathcal{K}} \prod _{i=1}^N\l_i^{\a-1}
\prod_{1\leq i<j\leq N}(\l_i-\l_j)^2\ \,d\L\,d\g.
\end{eqnarray} Therefore, formulas (\ref{Hilbert}) and
(\ref{Estimation-h:geq1}) imply
\begin{equation*}V_{\a}(\mathcal{K})\leq N^{(1-\a)N} V_{HS}(\mathcal{K}), \ \ for \ \
\a>1. \end{equation*}

\vskip 3mm  Now let us prove the lower bound for $\a>1$. By
(\ref{avolume:K})\begin{eqnarray} V_{\a}(\mathcal{K})&=&\int
_{\mathcal{K}} \prod _{i=1}^N\l_i^{\a-1} \prod_{1\leq i<j\leq
N}(\l_i-\l_j)^2 \,d\L\,d\g =\int _{\cK}f\ g \ \,d\L \,d\g,
\label{avolume:f:g}
\end{eqnarray}
where, to reduce the clutter, we denoted \begin{eqnarray*} &&g(\l
_1, \cdots, \l _N) = \prod _{i=1}^N
\l_i^{{\a-1}} \prod _{1\leq i<j\leq N}|\lambda_i-\lambda_j|^{2-2p}, \nonumber \\
&& f(\l _1, \cdots, \l _N)=\prod _{1\leq i<j\leq
N}|\lambda_i-\lambda_j|^{2p}.\end{eqnarray*} For any $p>1$, we
employ the H\"{o}lder inequality (see \cite{HLP}) to
(\ref{avolume:f:g}) and get
\begin{eqnarray}\label{avolume:f:g:ageq1}
V_{\a}(\mathcal{K})\! \geq \!\bigg(\!\int _{\cK}\! f^{\frac{1}{p}}
\,d\L \,d\g \bigg)^{p} \bigg(\int _{\cK} g^{\frac{1}{1-p}}\,d\L
\,d\g \bigg)^{1-p}.
\end{eqnarray}
By formula (\ref{Hilbert}), and substituting $f$ into the first
integral of (\ref{avolume:f:g:ageq1}), one gets
\begin{eqnarray}\label{Hilbert:K} \bigg(\int _{\cK} \!\prod _{1\leq i<j\leq N }(\lambda_i-\lambda_j)^2\,d\L \,d\g
\bigg)^{p}\!\!=\!\!V_{HS}(\mathcal{K})^{p}.
\end{eqnarray}
\par \noindent By substituting $g$ into the second integral of (\ref{avolume:f:g:ageq1}), one
has, for $p>\a>1$,
\begin{eqnarray}& &\bigg(\int _{\cK}\prod_{1\leq
i<j\leq N}|\lambda_i-\lambda_j|^{2} \prod _{i=1}^N \lambda _i
^{(\frac{\a-p}{1-p}-1)} \,d\L \,d\g\bigg)^{1-p} \nonumber \\
& &\geq \bigg(\int _{\cD}\prod_{1\leq i<j\leq
N}|\lambda_i-\lambda_j|^{2}\prod _{i=1}^N \lambda _i
^{(\frac{\a-p}{1-p}-1)}\,d\L \,d\g \bigg)^{1-p}=
\big(V_{\frac{\a-p}{1-p}}(\cD)\big)^{1-p},
\label{avolume:f:g:ageq1:1}
\end{eqnarray}
where the inequality follows $\cK\subset \cD$ and $p>\a>1$. The
lower bound is then an immediate consequence of inequalities
(\ref{Hilbert:K}) and (\ref{avolume:f:g:ageq1:1}).


\vskip 3mm \par \noindent {\bf Remark.} More generally, one can
compare $V_{\a}(\cK)$ in terms of $V_{\beta}(\cK)$, for all $\a,
\beta>0$. The above approach gives the following corollary.

\bc \label{avolume:bvolume:lemma:2} Let $\a, \beta>0$ be (fixed)
constants independent of the dimension of $\cH$. Let $\cH$ be any
(complex) Hilbert space with (complex) dimension $N$ and $\cK$ be
any measurable subset in $\cD(\cH)$. \vskip 3mm \noindent  (i):
For all $p>\frac{\a}{\beta}>1$,\begin{eqnarray*}
{V_{\beta}(\mathcal{K})}^{p}\bigg(V_{\frac{\a-\beta p}{1-p}}
(\cD) \bigg)^{1-p} \leq V_{\a}(\mathcal{K}) \leq N^{(\beta-\a)N} 
V_{\beta}(\mathcal{K}),
\end{eqnarray*}

\noindent (ii): For all $0<p<\frac{\a}{\beta}<1$,\begin{eqnarray*}
N^{(\beta-\a)N}
V_{\beta}(\mathcal{K}) \leq V_{\a}(\mathcal{K})\leq
{V_{\beta}(\mathcal{K})}^{p} \bigg(V_{\frac{\a-\beta p}{1-p}}
(\cD) \bigg)^{1-p}.
\end{eqnarray*}
\ec

The following lemma is simple but important for the proof of our
main results. \bl \label{a:hs} Let $\a>0$ be a constant
independent of $N$. Then
$$\lim _{N\rightarrow \infty} \bigg(\frac{V_{HS}(\cD)}{V_{\a}(\cD)}\bigg)^{\frac{1}{N^2-1}}=1.$$ \el
\vskip 2mm \noindent{\bf Remark.} A direct consequence of this
lemma is: let $\a, \beta>0$ be two constants independent of $N$,
then $$\lim _{N\rightarrow \infty}
\left(\frac{V_{\beta}(\cD)}{V_{\a}(\cD)}\right)^{\frac{1}{N^2-1}}=1.$$

\vskip 3mm \noindent {\bf Proof.} Now by formula
(\ref{avolume:D}),  one gets
\begin{equation}\label{aVShs}\frac{V_{HS}(\cD)}{V_{\a}(\cD)}=\frac{\G(N^2-N+\a
N)}{\G(N^2)}\ \prod _{j=1}^N
\bigg(\frac{\G(j)}{\G(j+\a-1)}\bigg).\end{equation} For each $j=1,
\cdots, N$, and by $\a >1$, one has
$$\frac{\G(j)}{\G(\a+j-1)}=\frac{1\times 2\times\cdots (j-1)}{\G (\a)\times \a \times (\a+1)\times \cdots  (\a+j-2)}\leq \frac{1}{\G(\a)}.$$
Combining with Stirling approximation formula (\ref{Stir:1}), one
has, for $\a>1$
$$\lim _{N\rightarrow \infty}\left(\frac{
V_{HS}(\mathcal{D})}{V_{\a}(\mathcal{D})}\right)^{1/(N^2-1)}\leq
1.$$

\vskip 3mm  Now we bound
$\left(\frac{V_{HS}(\cD)}{V_{\a}(\cD)}\right)^{{1}/{d}} $ from
below.  By induction, for all $j=1, \cdots, N$, and by $\a
>1$, one has
$$\frac{\G(j)}{\G(j+\a-1)}\geq \frac{\G(N)}{\G(N+\a-1)},$$  and
hence \begin{equation}\label{prod:lower}\prod
_{j=1}^N\bigg(\frac{\G(j)}{\G(j+\a-1)}\bigg)\geq
\bigg(\frac{\G(N)}{\G(N+\a-1)}\bigg)^N.\end{equation} Together
with Stirling approximation formula (\ref{Stir:1}) and formula
(\ref{aVShs}), one
$$\lim _{N\rightarrow \infty}
\bigg(\frac{V_{HS}(\cD)}{V_{\a}(\cD)}\bigg)^{\frac{1}{N^2-1}}\geq
1,
$$ which completes the proof.

\vskip 3mm  \bt\label{avolume:Hilbert:Theorem:1} Let $\a>0$ be a
(fixed) constant independent of the dimension of $\cH$. There are
universal constants (independent of dimension $N$, but depending
on $\a$) $C_1, c'_1>0$, such that for any Hilbert space $\cH$ and
any subset $\cK\subset \cD$,\vskip 3mm \noindent (i): for $\a>1$,
$ \mathrm{VR}_{\a}(\cK, \cD) \leq C_1 \ \mathrm{{VR}}_{HS}(\cK,
\cD), $
\\
(ii): for $0<\a<1$, $ \mathrm{VR}_{\a}(\cK, \cD)\geq  c'_1 \
\mathrm{{VR}}_{HS}(\cK, \cD). $

\vskip 2mm \noindent Moreover, the bounds are optimal in general.
\et

\vskip 2mm \noindent {\bf Remark.} In fact, as $N\rightarrow
\infty$, $C_1 (N,\a)\rightarrow 1$ (and $c_1'(N,\a)\rightarrow
1$). For small dimension $N$, one can precisely calculate
$C_1(N,\a)$ (and $c_1'(N,\a)$) by formula
(\ref{precise:constant:1}), if $\a$ is given. In fact, one can let
$$C_1(N,\a)=\left(\frac{N^{(1-\a)N}\
V_{HS}(\mathcal{D})}{V_{\a}(\mathcal{D})}\right)^{1/(N^2-1)}.$$
Moreover, by Lemma \ref{avolume:Hilbert:lemma:1}, $C_1(N,\a)\geq
1$ (and $c_1'(N,\a)\leq 1$) for all $\a>0$ and all $N$. Let us
consider $\cH=({\bbC^2})^{\otimes n}$ with $N=2^n$. Then, if
$\a=2$, $C_1(4,2)\approx 1.19861$, $C_1(8,2)\approx 1.10785$,
$C_1(16,2)\approx 1.05706$, $C_1(32,2)\approx 1.02958$ which seems
to decrease to $1$ as $N$ increases; while if $\a=1/2$,
$c'_1(4,1/2)\approx 0.873608$, $c'_1(8,1/2)\approx 0.936271$,
$c'_1(16,1/2)\approx 0.968159$, $c'_1(32,1/2)\approx 0.984148$
which seems to increase to $1$ if $N$ increases. Similar
phenomenon happens to other fixed $\a$, and this strongly suggests
that $C_1 (4, \a)$ for $\a>1$ or $c'_1(4, \a)$ for $\a<1$ should
work for all $2^n\geq 4$ (and, in fact, all other integers $N\geq
4$).

\vskip 3mm \noindent {\bf Proof.} We only prove the case of
$\a>1$. Now dividing $V_{\a}(\cD)$ from both sides of the upper
bound of inequality (\ref{avolume:Hilbert:ageq1}), then
$$\frac{V_{\a}(\mathcal{K})}{V_{\a}(\mathcal{D})} \leq \frac{N^{(1-\a)N}\ V_{HS}(\mathcal{K})}{V_{\a}(\mathcal{D})}
= \left(\frac{V_{HS}(\cK)}{V_{HS}(\cD)}\right)\
\left(\frac{N^{(1-\a)N}\
V_{HS}(\mathcal{D})}{V_{\a}(\mathcal{D})}\right).$$ Taking $d$-th
root from both sides, then
\begin{equation}\label{precise:constant:1} \mathrm{VR}_{\a}(\cK,\cD) \leq
\mathrm{VR}_{HS}(\cK, \cD )\ \left(\frac{N^{(1-\a)N}\
V_{HS}(\mathcal{D})}{V_{\a}(\mathcal{D})}\right)^{1/(N^2-1)}.\end{equation}
By Lemma \ref{a:hs}, for $\a>1$,
$$\lim _{N\rightarrow \infty}\left(\frac{N^{(1-\a)N}\
V_{HS}(\mathcal{D})}{V_{\a}(\mathcal{D})}\right)^{1/(N^2-1)}= 1,$$
and hence, there is a universal constant $C_1>0$ such that
$\mathrm{VR}_{\a}(\cK, \cD)\leq C_1 \mathrm{VR}_{HS}(\cK, \cD).$

\vskip 2mm To see the optimality of the upper bound, we let
$\cK_t=t \cD +(1-t )\r _{max}$ for $0<t<1$, i.e.,
$$\cK_t\!=\!\left\{\! UXU^\dag\!:\! X\!=\!{\rm diag}\!\left(\!\frac{1-t}{N}+t\lambda_1,
\cdots, \frac{1-t}{N}+t\lambda_N\!\right),\! \mbox{$(\l _1,\cdots
\l_N)\in \D$ and $U\in \cU(N)$}\! \right\}.$$  Let $Z_N$ be as in
(\ref{gam-measure}).

\vskip 3mm Clearly $\mathrm{VR}_{HS}(\cK _t,\cD)=t$ by homogeneity
of the Hilbert-Schmidt volume. Now we estimate $V_{\a}(\cK_t)$
from below. In fact
$$V_{\a}(\cK_t)=Z_N t^{N^2-1}\int _{\D _1} \prod _{j=1}^N
\bigg(\frac{1-t}{N}+t\lambda_j\bigg)^{\a-1} \prod_{1\leq i<j\leq
N}(\l_i-\l_j)^2\,\prod_{j=1}^{N-1}\,d\l_j.
$$
As $\frac{1-t}{N}+t\lambda_j\geq t\l_j$ for all $j=1, \cdots, N$,
and $\a>1$, one has $$V_{\a}(\cK_t)\!\geq\! \! t^{N^2-1+N(\a-1)}
Z_N \!\int _{\D _1}\!\! \prod _{j=1}^N \lambda_j^{\a-1}
\!\!\prod_{1\leq i<j\leq
N}(\l_i-\l_j)^2\!\!\prod_{j=1}^{N-1}\!\!d\l_j= t^{N^2-1+N(\a-1)}
V_{\a}(\cD).$$ Equivalently, $$\mathrm{VR}_{\a}(\cK_t, \cD)\geq t
\cdot t^{\frac{N(\a-1)}{N^2-1}}.$$ Hence, if $t\geq
exp(-c_1''(N^2-1)/N)$ for some constant $c_1''>0$, then
$$\mathrm{VR}_{\a}(\cK _t, \cD)\geq exp(-c_1''(\a-1))t=exp(-c_1''(\a-1))\mathrm{VR}_{HS}(\cK _t,\cD). $$
Part (i) guarantees that $\mathrm{VR}_{\a}(\cK _t, \cD)\leq C_1
\mathrm{VR}_{HS}(\cK _t,\cD).$ So the upper bound can be obtained
for $\a>1$, and it is optimal in general.


\vskip 3mm \bt \label{avolume:Hilbert:Theorem:2} Let $\a>0$ be a
(fixed) constant independent of the dimension of $\cH$. There are
universal constants (independent of $N$, but depending on $\a$)
$c_1, C'_1>0$, such that for any Hilbert space $\cH$ and any
subset $\cK\subset \cD$ (let $\zeta=\mathrm{VR}_{HS}(\cK, \cD)$),
\vskip 3mm \noindent (i): for $\a>1$, $\mathrm{VR}_{\a}(\cK, \cD)
\geq c_1 \ \mathrm{{VR}}_{HS}^{\a}(\cK,
\cD)\exp\bigg(\!\frac{(1-\a)\ln \ln (e/\zeta)}{N^2-1}\!\bigg), $
\\
(ii): for $0<\a<1$, $ \mathrm{VR}_{\a}(\cK, \cD)\leq C'_1 \
\mathrm{{VR}}_{HS}^{\a}(\cK, \cD)\exp\bigg(\!\frac{(1-\a)\ln \ln
(e/\zeta)}{N^2-1}\!\bigg).$   \et

\vskip 2mm \noindent {\bf Proof.} We only prove (i). By the lower
bound of (\ref{avolume:Hilbert:ageq1}), for all $p>\a>1$,
$${V_{HS}(\mathcal{K})}^{p}\ \bigg(V_{\big(\frac{\a-p}{1-p}\big)}
(\cD) \bigg)^{1-p} \leq V_{\a}(\mathcal{K}). $$ Dividing
$V_{\a}(\cD)$ from both sides, one has for all $p>\a>1$,
\begin{eqnarray*}\frac{V_{\a}(\mathcal{K})}{V_{\a}(\cD)} &\geq&
\bigg(\frac{V_{HS}(\mathcal{K})}{V_{HS}(\cD)}\bigg)^{p} \
\bigg(\frac{V_{HS}(\cD)}{V_{\a}(\cD)}\bigg)^p \ \bigg(
\frac{V_{(\frac{\a-p}{1-p})}(\cD)}{V_{\a}(\cD)} \bigg)^{1-p}.
\end{eqnarray*} Equivalently, by taking the $d$-th root from both
sides,
\begin{eqnarray}\mathrm{VR}_{\a}(\mathcal{K},\cD) &\geq&
\big(\mathrm{VR}_{HS}(\mathcal{K}, \cD)\big)^{p} \
\bigg(\frac{V_{HS}(\cD)}{V_{\a}(\cD)}\bigg)^{\frac{p}{d}} \ \bigg(
\frac{V_{(\frac{\a-p}{1-p})}(\cD)}{V_{\a}(\cD)}
\bigg)^{\frac{1-p}{d}}. \label{alpha:lower}
\end{eqnarray}
We first bound $\left(
\frac{V_{(\frac{\a-p}{1-p})}(\cD)}{V_{\a}(\cD)}
\right)^{\frac{1-p}{d}}$ from below. To that end, we let $\t=\t
(N, \zeta )=\frac{p-\a}{p-1}=\frac{1}{N\ln (e/\zeta)}$ be a
function depending on $N$ and $\zeta=\mathrm{VR}_{HS}(\cK, \cD)$.
This implies that $p(N,\zeta)=\frac{\a-\t}{1-\t}$, and
$p(N,\zeta)\rightarrow \a$ as $N\rightarrow \infty$. Moreover,
$0<\t N=\frac{1}{\ln(e/\zeta)}<1$, and hence $\t j<1$ for all
$j=1, \cdots, N-1$.

\vskip 3mm \noindent  By formula (\ref{avolume:D}), one has
$$\frac{V_{\t}{(\cD)}}{V_{\a}(\cD)}=\frac{\G(N^2-N+\a N)}{\G(N^2-N+\t N)} \prod _{j=1}^N\bigg(\frac{\G(j+\t-1)}{\G(j+\a-1)}\bigg).$$
Formula (\ref{YD3}) and induction show that for all $j=2, \cdots,
N$,
$$\frac{\G(j+\t-1)}{\G(j+\a-1)}
\leq \frac{\G(\t+1)}{\G(\a+1)}\leq \frac{1}{\G(\a+1)},$$ because
$0<\t<1$. Thus, again by (\ref{YD3})
$$\prod_{j=1}^N\bigg(\frac{\G(j+\t-1)}{\G(j+\a-1)}\bigg) \leq \bigg(\frac{1}{\G(\a+1)}\bigg)^{N-1} \
\frac{\G(\t)}{\G(\a)}\leq \frac{\a}{\t\ [\Gamma(1+\a)]^N}.$$ This
further implies, by $p=\frac{\a-\t}{1-\t}>1$ and $\t=\frac{1}{N\ln
(e/\zeta)}$,
\begin{eqnarray}&&\lim_{N\rightarrow \infty}\bigg(\prod_{j=1}^N\frac{\G(j+\t-1)}{\G(j+\a-1)}\bigg)^{\frac{1-p}{N^2-1}}
\geq \lim_{N\rightarrow \infty} \left(\frac{\a}{\t\
[\Gamma(1+\a)]^N}\right)^{\frac{1-p}{N^2-1}}\nonumber
\\&& \ \ \geq \!\! \lim _{N\rightarrow \infty}\!
\left(\!\frac{\a}{[\Gamma (1+\a)]^N}\!\right)^{\frac{1-p}{N^2-1}}
\exp\! \bigg(\!\big(\frac{1-\a}{N^2-1}\big) (\ln N+1)\!\bigg)\!
\exp\bigg(\!\frac{(1-\a)\ln \ln (e/\zeta)}{N^2-1}\!\bigg)\nonumber\\
&&\ \ =\lim _{N\rightarrow \infty}\exp\bigg(\frac{(1-\a)\ln \ln
(e/\zeta)}{N^2-1}\bigg).\label{a:tao}\end{eqnarray} Stirling
approximation formula (\ref{Stir:1}) implies that
$$\lim_{N\rightarrow \infty} \bigg(\frac{\G(N^2-N+\a N)}{\G(N^2-N+\t
N)}\bigg)^{\frac{1-p}{N^2-1}}=1.$$ Together with inequality
(\ref{a:tao}), one has
\begin{equation}\label{a:tao:1}\lim_{N\rightarrow
\infty}\bigg(\frac{V_{\t}(\cD)}{V_{\a}(\cD)}\bigg)^{\frac{1-p}{N^2-1}}\geq
\lim _{N\rightarrow \infty} \exp\bigg(\frac{(1-\a)\ln \ln
(e/\zeta)}{N^2-1}\bigg).\end{equation} Inequality
(\ref{avolume:Hilbert:ageq1}) implies that
\begin{equation}\label{a:HS:1}\lim_{N\rightarrow
\infty}\left(\frac{V_{HS}(\cD)}{V_{\a}(\cD)}\right)^{\frac{p}{d}}\geq
\lim_{N\rightarrow \infty} N^{\frac{p(\a-1)N}{N^2-1}}\geq
\lim_{N\rightarrow \infty} \exp\bigg(\frac{\a(\a-1)\log N}
{N}\bigg)=1.\end{equation} Since $\t \ln\zeta=\frac{\ln
\zeta}{N(1+\ln(1/\zeta))}\geq \frac{-1}{N}$ and $0<\t<1$, one has
\begin{equation}\label{vrhs:a:p} \lim _{N\rightarrow \infty}\mathrm{VR}_{HS}(\cK,\cD)^{p-\a}=\lim _{N\rightarrow \infty} exp\bigg(\frac{\a-1}{1-\t}\ \t
\ln\zeta\bigg)\geq \lim _{N\rightarrow \infty}
exp\bigg(\frac{(1-\a)}{N}\bigg)=1.\end{equation} Combining
(\ref{alpha:lower}), (\ref{a:tao:1}), (\ref{a:HS:1}), and
(\ref{vrhs:a:p}), one gets $\mathrm{VR}_{\a}(\cK, \cD)\! \geq\!
c_1 \mathrm{{VR}}_{HS}^{\a}(\cK, \cD)\exp\bigg(\!\frac{(1-\a)\ln
\ln (e/\zeta)}{N^2-1}\!\bigg)$ for some universal constant
$c_1>0$.

\vskip 3mm \noindent {\bf Remark.} A slightly more precise
calculation shows that $$\mathrm{VR}_{\a}(\cK, \cD)\! \geq\! c_1
\mathrm{{VR}}_{HS}^{\a}(\cK, \cD)\exp\bigg(\!\frac{(1-\a)\ln \ln
(e/\zeta)}{N^2-1}\!\bigg)\bigg(1+O\left(\frac{\ln N
}{N}\right)\bigg).$$ Moreover, from the above proof, one gets for
any fixed $\a>1$, the constant $c_1(N,\a)\rightarrow 1$ as
$N\rightarrow \infty$. Numerical results strongly suggest that,
for any fixed $\a>1$, $c_1(N,\a)$ increases to $1$ as $N$
increases, and $c_1(4, \a)$ should work for all $N\geq 4$. For
instance, if we consider $\cH=(\bbC^2)^{\otimes n}$ with $N=2^n$,
then our proof yields that $c_1(4, 2)\approx 0.219056$, $c_1(8,
2)\approx 0.374655$, $c_1(16, 2)\approx 0.539382$, $c_1(32,
2)\approx 0.686509,$ $c_1(64, 2)\approx 0.800189$, $c_1(128,
2)\approx 0.878831,$ $c_1(256, 2)\approx 0.929124$, and $c_1(512,
2)\approx 0.959595$. Similar phenomenon happens to the case of
$0<\a<1$, namely $C_1'(N,\a)$ decrease to $1$ as $N$ increases.
For example, if $\a=0.5$, then $C_1'(4,0.5)\approx 1.43603$,
$C_1'(8,0.5)\approx 1.24359$, $C_1'(16,0.5)\approx 1.13648$,
$C_1'(32,0.5)\approx 1.07664$, $C_1'(64,0.5)\approx 1.04292$, and
hence, $C_1'(4,\a)$ should work for all $N\geq 4$.

 \vskip 3mm As a consequence of Corollary
\ref{avolume:bvolume:lemma:2}, one can prove the following result,
whose proof is similar to Theorems \ref{avolume:Hilbert:Theorem:1}
and \ref{avolume:Hilbert:Theorem:2}.

\bc \label{avolume:bvolume:lemma:3} Let $\a, \beta\!\!>\!\!0$ be
(fixed) constants independent of the dimension of $\cH$. Let $\cH$
be any (complex) Hilbert space with (complex) dimension $N$ and
$\cK$ be any measurable subset in $\cD(\cH)$. Let
$\zeta\!=\!{\mathrm{VR}_{\beta}(\mathcal{K},\cD)}$,
$\a_{\max}\!=\!\max\{1,\frac{\a}{\beta}\}$, and
$\a_{\min}\!=\!\min\{1,\frac{\a}{\beta}\}$. There exist universal
constants (independent of $N$, but depending on $\a$ and $\beta$)
$c_2, C_2\!>\!0$, such that,
$$ c_2\
\zeta^{\a _{\max} }\ \exp\!\left(\!\frac{(1-\a_{\max}) \ln\ln
(e/\zeta)}{N^2-1}\!\right)\! \leq\! \mathrm{VR}_{\a}(\mathcal{K},
\cD)\! \leq\! C_2\ \zeta^{\a_{\min}}\
\exp\!\left(\!\frac{(1-\a_{\min})\ln\ln
(e/\zeta)}{N^2-1}\!\right).
$$ \ec

\noindent {\bf Remark.} As in Theorem
\ref{avolume:Hilbert:Theorem:1}, $\cK_t$ attains the upper bound
if $\a>\beta$, and attains the lower bound if $\a<\beta$.
Therefore, for $\a>\beta$ the upper bound is optimal in general,
and for $\a<\beta$ the lower bound is optimal in general. Theorems
\ref{avolume:Hilbert:Theorem:1} and
\ref{avolume:Hilbert:Theorem:2} are special cases of this
corollary with $\beta=1$.


\vskip 3mm Theorems 1 and 2 in \cite{YeDeping2009} compare the
Bures volume of $\cK$ in terms of the Hilbert-Schmidt volume. The
following theorem compares the Bures volume with the $\a$-volume.

\bt\label{Bures:a:compare} Let $\a>0$ be a (fixed) constant
independent of the dimension of $\cH$. Let $\cH$ be any (complex)
Hilbert space with (complex) dimension $N$ and $\cK$ be any
measurable subset in $\cD(\cH)$. There exist universal constants
(independent of $N$, but depending on $\a$) $c_3, C_3>0$, such
that
$$c_3 \ \xi^{\max\{1, \frac{1}{2\a}\}}\!\exp\!\left(\!\frac{(1\!-\!\max\{1,
\frac{1}{2\a}\})\!\ln\ln(e/\xi)}{N^2-1}\!\right)\!\leq\!
\mathrm{VR}_{B}(\cK, \cD)\leq C_3 \ \xi^{\min\{\frac{1}{2},
\frac{1}{2\a}\}}\exp\left(\frac{\ln \ln (e/ \xi)}{2N}\right),$$
where $\xi=\mathrm{VR}_{\a}(\cK, \cD)$.
 \et

 \vskip 3mm \noindent {\bf Proof.} Similar to the proof of Lemma
 \ref{avolume:Hilbert:lemma:1}, $\prod _{1\leq i<j\leq N}
 \frac{1}{\l_i+\l_j}$ attains the minimum at $\l_i=1/N$ for all
 $i=1, \cdots, N$. Hence,
 \begin{eqnarray*}
V_B(\cK)&=&\int _{\cK} \frac{1}{\sqrt{\l_1 \cdots \l_N}} \prod
_{1\leq i<j\leq N}
 \frac{(\l_i-\l_j)^2}{\l_i+\l_j}\,d\L\,d\g \\&\geq&
 \left(\frac{N}{2}\right)^{\frac{N^2-N}{2}} \int _{\cK} \frac{1}{\sqrt{\l_1 \cdots \l_N}} \prod
_{1\leq i<j\leq N}
 (\l_i-\l_j)^2\,d\L\,d\g \\&=&\left(\frac{N}{2}\right)^{\frac{N^2-N}{2}}
 V_{\frac{1}{2}} (\cK).
 \end{eqnarray*}
Recall $d=N^2-1$. Dividing $V_{B}(\cD)$ from both sides and taking
the $d$-th root, one gets that for some universal constant
$c''>0$,
\begin{eqnarray}
\mathrm{VR}_B(\cK, \cD)
&\geq&\left(\frac{N}{2}\right)^{\frac{N^2-N}{2(N^2-1)}}
 \mathrm{VR}_{\frac{1}{2}} (\cK,\cD)
 \left(\frac{V_{\frac{1}{2}}(\cD)}{V_{B}(\cD)}\right)^{1/d}\nonumber
 \\ &\geq & c'' \mathrm{VR}_{\frac{1}{2}}(\cK, \cD).\label{Bures:1/2}
 \end{eqnarray}
where the second inequality follows Lemma \ref{a:hs} and the limit
\cite{YeDeping2009}
$$\lim_{N\rightarrow \infty}N^{\frac{N^2-N}{2(N^2-1)}}
 \left(\frac{V_{HS}(\cD)}{V_{B}(\cD)}\right)^{1/d}=2e^{-1/4}.$$

 \vskip 1mm The lower bound is an immediate consequence of
 Corollary \ref{avolume:bvolume:lemma:3} and inequality
 (\ref{Bures:1/2}). The upper bound is an immediate consequence of
 Theorems \ref{avolume:Hilbert:Theorem:1},
 \ref{avolume:Hilbert:Theorem:2} and Theorem 2 in
 \cite{YeDeping2009}.

\vskip 3mm \noindent {\bf Remark.} If $\a\geq \frac{1}{2}$, the
lower bound is optimal in general, and if $\a\leq 1$, the upper
bound is optimal in general. If $\a\in [1/2,1]$, then there exist
universal constants $c_3', C_3'>0$, such that
$$c_3' \mathrm{VR}_{\a}(\cK, \cD)\leq
\mathrm{VR}_{B}(\cK, \cD)\leq C_3' \ [\mathrm{VR}_{\a}(\cK,
\cD)]^{1/2}\exp\left(\frac{\ln \ln (e/ \xi)}{2N}\right),$$ and the
bounds are optimal in general. Moreover, there exist universal
constants $c_3'', C_3''>0$, such that
$$c_3'' \mathrm{VR}_{\frac{1}{2}}(\cK, \cD)\leq
\mathrm{VR}_{B}(\cK, \cD)\leq C_3'' \ [\mathrm{VR}_{HS}(\cK,
\cD)]^{1/2}\exp\left(\frac{\ln \ln (e/ \zeta)}{2N}\right),$$ with
$\zeta=\mathrm{VR}_{HS}(\cK,\cD)$ and the bounds are optimal in
general. This inequality improves results in \cite{YeDeping2009}.

 \vskip 3mm In fact, it was proved that $\bar{c}t\leq \mathrm{VR}_B(\cK_t, \cD)\leq \bar{C}t
 $ if $t\leq \frac{3}{4}$ for some universal constants $\bar{c},
 \bar{C}>0$  \cite{YeDeping2009}. Recall $\cK_t=t \cD +(1-t )\r _{max}$. Therefore,
 $\cK_t$ attains the lower bound for $\a\geq 1/2$ by a similar
 calculation in the proof of Theorem
 \ref{avolume:Hilbert:Theorem:1}.

 \vskip 3mm  To see the optimality of the upper bound for $\a\leq
 1$, we let $0<t<1$, and
$$\!\cK^t\!=\!\{UXU^\dag: X=\!{\rm diag}(1-t+t\lambda_1, t\lambda_2, \cdots,
t\lambda_N), \! \ \mbox{$(\l _1,\cdots \l_N)\in \D_1$ and $U\in
\cU(N)$} \!\}.$$ Recall $\D _1$ is the chamber of $\D$ with order
$\l_1 \geq \cdots \geq \l _N$. For
$e^{\tilde{c}(-1-N)}<t<\frac{4}{5}$, one has
$\mathrm{VR}_{B}(\cK^t, \cD)\geq
\hat{c}\sqrt{\mathrm{VR}_{HS}(\cK^t, \cD)}$ where $\tilde{c},
\hat{c}>0$ are two constants \cite{YeDeping2009}. To prove the
optimality of the upper bound for $\a\leq 1$, it is enough to
prove that $\mathrm{VR}_{HS}(\cK^t,\cD)\geq \check{c}
\mathrm{VR}_{\a}(\cK^t,\cD)$ for some constant $\check{c}>0$.

\vskip 3mm To that end,  we first bound $V_{\a}(\cK^t)$ from
above. By formula (\ref{avolume:element-1}), one has
\begin{eqnarray*} V_{\a}(\!\cK^t)\!&=&\!Z_N
t^{(N-1)^2+(\a-1)(N-1)}\! \int _{\D_1}\!\!\prod_{j=2}^N
\!\l_j^{\a-1}\! \prod_{2\leq i<j\leq N}\big(\l_i-\l_j\big)^2 \\ &
& \times \!(1+t\l
_1-t)^{\a-1}\!\prod_{j=2}^N(1-t+t\l_1-t\l_j)^2\!\prod
_{j=1}^{N-1}\,d\l_j.
\end{eqnarray*}
As $1-t\leq 1+t\l _1-t\leq 1$, $0\leq 1-t+t\l_1-t\l_j\leq 1$, and
$\a\leq 1$, one has
\begin{eqnarray*} V_{\a}(\!\cK^t)\!&\leq&\!Z_N
t^{(N-1)^2+(\a-1)(N-1)}\! (1-t)^{\a-1}\! \int
_{\D_1}\!\!\prod_{j=2}^N \!\l_j^{\a-1}\! \prod_{2\leq i<j\leq
N}\big(\l_i-\l_j\big)^2 \!\prod _{j=1}^{N-1}\,d\l_j\\ &=&
\frac{Z_N}{Z_{N-1}} t^{(N-1)^2+(\a-1)(N-1)}\! (1-t)^{\a-1}\!
V_{\a}(\cD_{N-1}) \int _{0}^1
(1-\l_1)^{N^2-2N+(N-1)(\a-1)}\,d\l_1\\ &\leq & \frac{Z_N}{Z_{N-1}}
t^{(N-1)^2+(\a-1)(N-1)}\! (1-t)^{\a-1}\! V_{\a}(\cD_{N-1}).
\end{eqnarray*}

\vskip 2mm \noindent Stirling approximation formula (\ref{Stir:1})
implies that, for $e^{\tilde{c}(-1-N)}<t<\frac{4}{5}$,
\begin{equation}\label{K:T}\mathrm{VR}_{\a}(\cK^t, \cD) \leq \hat{C} t,
\end{equation} for some constant $\hat{C}>0$.

\vskip 3mm Now we bound $V_{HS}(\cK^t)$ from below. That is,
\begin{eqnarray*} V_{HS}(\!\cK^t)\!&=&\!Z_N
t^{(N-1)^2}\! \int _{\D_1}\!\ \prod_{2\leq i<j\leq
N}\big(\l_i-\l_j\big)^2 \!\prod_{j=2}^N(1-t+t\l_1-t\l_j)^2\!\prod
_{j=1}^{N-1}\,d\l_j\\ &\geq & Z_N t^{N^2-1} V_{HS}(\cD),
\end{eqnarray*} where the inequality follows $1-t+t\l_1-t\l_j\geq t\l_1-t\l_j$ for all $j=2, \cdots, N$.
 Equivalently, $\mathrm{VR}_{HS}(\cK^t, \cD)\geq t$.
Together with inequality (\ref{K:T}), one has
$$\mathrm{VR}_{HS}(\cK^t,\cD)\geq \check{c}
\mathrm{VR}_{\a}(\cK^t,\cD), $$ for
$\check{c}=\frac{1}{\hat{C}}>0$. This concludes the optimality of
the upper bound for $\a\leq 1$.

\section{the $\a$-volumes of $\cS$ and $\mathcal{PPT}$}
In this section, we provide answers to the questions (1) and (3)
for large $N$ in the sense of $\a$-volume. These two questions are
related to the ones listed on page 9 in \cite{K. Zyczkowski1998},
that is,
\begin{itemize}
\item Does the volume of the set of separable states go really to zero as the
dimension of the composite system $N$ grows, and how fast?
\item Has the set of separable states really a volume strictly
smaller than the volume of the set of states with a positive
partial transpose?
\end{itemize} Here, we supply solutions to those two
questions in the sense of $\a$-volume for large $N$.

 \vskip 2mm The geometry of $\cD$ is reasonably well-understood,
and hence the volumes (e.g., Hilbert-Schmidt, Bures, $\a$) of
$\cD$ can be calculated precisely \cite{Zycz2, Zycz1, Ben1, Soz1}.
However, the geometry of $\cS$ and $\mathcal{PPT}$ are more
complicate, and less information for them is available. This makes
the calculation of volumes (Hilbert-Schmidt, Bures, $\a$) of $\cS$
and $\mathcal{PPT}$ quite intractable. Thanks to the remarkable
works of \cite{AS1, Sz1}, one knows that $\mathrm{VR}_{HS}(\cK,
\cD)$ decays to $0$ very quick for large $N$ and
$\mathrm{VR}_{HS}(\mathcal{PPT}, \cD)$ is essentially a constant.
In \cite{YeDeping2009}, the author obtained similar results in the
sense of Bures volume. Therefore, the probability of finding
separable quantum states within quantum states is extremely small
and the Peres-Horodecki PPT criterion as tools to detect
separability is imprecise for large $N$, in the sense of both
Hilbert-Schmidt and Bures volumes.

\vskip 3mm The following two corollaries give the estimates of
$\a$-volume of $\cS$. They are direct consequences of Theorems
\ref{avolume:Hilbert:Theorem:1} and
\ref{avolume:Hilbert:Theorem:2}, and the estimates for
$\mbox{VR}_{HS}(\cS, \cD)$ implicit in \cite{AS1}.

\bc \label{Bures:S:1}{\bf (Large number of small subsystems)} Let
$\a>0$ be a (fixed) constant independent of the dimension of
$\cH$. For system $\cH=(\bbC^D)^{\otimes n}$, there exist
universal constants $c_4, C_4>0$, such that for all $D,n\geq 2$,
\begin{equation*} \label{YD1} \left(\frac{c_4}{N^{{1}/{2}+\a
_D}}\right)^{\max\{1,\a\}} \leq \mathrm{VR}_{\a}(\cS, \cD) \leq
C_4 \ \left( {\frac{(D n\ln n)^{1/2}}{N^{1/2+\a
_D}}}\right)^{\min\{1, \a\}},
\end{equation*}
where $\a _D =\frac{1}{2}\log _D
(1+\frac{1}{D})-\frac{1}{2D^2}\log _D(D+1)$.\ec

\vskip 3mm \noindent {\bf Remark.} Recall that for
$\cH=(\bbC^D)^{\otimes n}$, $N=D^n$. Corollary \ref{Bures:S:1}
means that, for fixed (small) $D$ and fixed $\a>0$,
$\mathrm{VR}_{\a}(\cS, \cD)$ is bounded above by (up to a
multiplicative universal constant) $\left({\frac{\sqrt{D n\ln
n}}{D^{n(1/2+\a _D)}}}\right)^{\min \{1,\a\}}$, and hence
$\mathrm{VR}_{\a}(\cS, \cD)$ goes to zero exponentially as
$n\rightarrow \infty$. This implies that $\bbP_{\a}(\cS, N,
D)=:\frac{V_{\a}(\cS)}{V_{\a}(\cD)}$, the $\a$-probability of
finding separable states within $\cD$ (on $\cH=(\bbC^D)^{\otimes
n}$), goes to zero super-double-exponentially, since
$\mathrm{VR}_{\a}(\cS, \cD) $ is already the $(N^2-1)$-th root of
$\bbP_{\a}(\cS, N, D)$. Numerical results show that, for even
(moderately) large $N$, $\bbP_{\a}(\cS, N, D)$ is very small. For
instance, $\bbP _{2}(\cS, 256, 2)\leq 2.1\times 10^{-1595}$ with
$n=8$, $\bbP _{2}(\cS, 512, 2)\leq 3.1\times 10^{-43631}$ with
$n=9$, $\bbP _{3}(\cS, 512, 2)\leq 1.1\times 10^{-43412}$ with
$n=9$, and $\bbP_{2}(\cS, 243, 3)\leq 1.52\times 10^{-5301}$ with
$n=5$. Similarly, we list some probability for $0<\a<1$, $\bbP
_{0.5}(\cS, 256, 2)\leq 8.8\times 10^{-479}$ with $n=8$, $\bbP
_{0.5}(\cS, 512, 2)\leq 3.36\times 10^{-21102}$ with $n=9$, $\bbP
_{0.2}(\cS, 512, 2)\leq 6.62\times 10^{-7905}$ with $n=9$, and
$\bbP _{0.5}(\cS, 243, 3)\leq 9.5\times 10^{-2351}$ with $n=5$.

\vskip 3mm \bc \label{Bures:S:2}{\bf (Small number of large
subsystems)} Let $\a>0$ be a (fixed) constant independent of the
dimension of $\cH$. For system $\cH=(\bbC^D)^{\otimes n}$, there
exist universal computable constants $c_5, C_5>0$, such that for
all $D,n\geq 2$,
\begin{equation*} \label{YD2} \left(\frac{c_5^n}{N^{1/2-1/(2n)}}\right)^{\max\{1,\a\}} \leq
\mathrm{VR}_{\a}(\cS, \cD) \leq C_5 \ \left({\frac{(n \ln
n)^{1/2}}{N^{1/2-1/(2n)}}}\right)^{\min\{1,\a\}}. \end{equation*}
\ec

\vskip 2mm \noindent {\bf Remark.} Corollary \ref{Bures:S:2} shows
that, for fixed (small) $n$, ``the order of decay" of
$\mathrm{VR}_{\a}(\cS,\cD)$ is between
$D^{(\frac{1}{2}-\frac{n}{2})\max\{1,\a\}}$ and
$D^{(\frac{1}{2}-\frac{n}{2})\min\{1,\a\}}$ as $D\rightarrow
\infty.$ Therefore, $\bbP_{\a}(\cS, N, D)$ goes to zero with order
between $D^{({1}/{2}-{n}/{2})(D^{2n}-1)\min\{1,\a\}}$ and
$D^{({1}/{2}-{n}/{2})(D^{2n}-1)\max\{1,\a\}}$ as $D\rightarrow
\infty.$ Numerical results show that $\bbP_{\a}(\cS, N, D)$ is
very small for (even moderately) large $N$, and its ``order of
decay" relies not only on $D$ but also (heavily) on $n$. For
instance, if $\a=2$, then $\bbP _{2}(\cS, 784, 28)\leq 1.35 \times
10^{-5230}$ with $n=2$, $\bbP _{2}(\cS, 900, 30)\leq 8.5 \times
10^{-19028}$ with $n=2$, $\bbP _{2}(\cS, 729, 9)\leq 4.9 \times
10^{-27218}$ with $n=3$, and $\bbP _{2}(\cS, 1000, 10)\leq
1.74\times 10^{-97132}$ with $n=3$. Similarly, if $\a=1/2$, then
$\bbP_{0.5}(\cS, 900, 30)\leq 1.12\times 10^{-8235}$ with $n=2$,
and $\bbP_{0.5}(\cS, 729, 9)\leq 6.8\times 10^{-12538}$ with
$n=3$.

\vskip 3mm For a bipartite system $\cH=\bbC ^{D_1} \otimes \bbC
^{D_2}$, any state $\rho $ on $\cH$ can be expressed uniquely as
$$\rho = \sum _{i,j}^{D_1} \sum _{\a, \beta }^{D_2} \rho _{i\a, j\beta } |e_i\otimes
f_{\a}\rangle \langle e_j\otimes f_{\beta}|$$ where
$\{e_i\}_{i=1}^{D_1}$ and $\{f_\a\}_{\a=1}^{D_2}$ are the
canonical bases of $\bbC ^{D_1}$ and $\bbC ^{D_2}$ respectively.
Define the partial transpose $T(\rho)$ with respect to the first
subsystem as
$$T(\r)=\sum _{i,j}^{D_1} \sum _{\a, \beta }^{D_2} \rho _{j\a, i\beta } |e_i\otimes
f_{\a}\rangle \langle e_j\otimes f_{\beta}|.$$  We use
$\mathcal{PPT}$ ($\subset \cD$) to denote the set of quantum
states with PPT, i.e., $\r\in \mathcal{PPT}$ if and only if
$T(\r)\geq 0$. Thus the Peres-Horodecki PPT criterion is
equivalent to $\cS\subset \mathcal{PPT}$. For systems $\bbC
^2\otimes \bbC^2$, $\bbC ^2\otimes \bbC^3$ and $\bbC ^3\otimes
\bbC^2$, one has $\cS=\mathcal{PPT}$. That is, the Peres-Horodecki
PPT criterion is a necessary and sufficient condition for $\r$
being a separable quantum state if dimension of $\cH$ less than or
equal to $6$.

\vskip 3mm The following corollary is to estimate the $\a$-volume
of $\mathcal{PPT}$. It is a direct consequence of Theorems
\ref{avolume:Hilbert:Theorem:1} and
\ref{avolume:Hilbert:Theorem:2}, and Theorem 4 in \cite{AS1}. Note
the upper bound is trivial because $\mathcal{PPT}\subset \cD$.

\bc \label{avolume:PPT} {\bf ($\a$-Volume of $\mathcal{PPT}$)} Let
$\a>0$ be a (fixed) constant independent of the dimension of
$\cH$. There exists an absolute computable constant (depending on
$\a$) $c_0>0$, such that, for any bipartite system $\cH=\bbC
^D\otimes \bbC ^D$, $c_0\leq \mathrm{VR}_{\a}(\mathcal{PPT},
\cD)\leq 1.$ \ec

\vskip 3mm \noindent {\bf Remark.} An immediate consequence of
Corollaries \ref{Bures:S:2} and \ref{avolume:PPT} is that, for
$\cH=\bbC ^D\otimes \bbC ^D$ and large $D$, ``the order of decay"
of $\mathrm{VR}_{\a} (\cS, \mathcal{PPT})$ is between
$D^{-\frac{1}{2}\max\{1,\a\}}$ and $D^{-\frac{1}{2}\min\{1,\a\}}$
as $D\rightarrow \infty.$ The upper bound decreases to $0$ as
$D\rightarrow \infty$, and hence $\cS$ has really a volume
strictly small than $\mathcal{PPT}$. Moreover, the conditional
$\a$-probability of separability given positive partial transpose
condition is exceedingly small. In other words, the
Peres-Horodecki PPT criterion is imprecise as tools to detect
separability for large $N$. We point out that our proof yields
$c_0(N, \a)\rightarrow (e^{-1/4}/4)^{\max\{1,\a\}}\approx
(0.195)^{\max\{1,\a\}}$ as $N\rightarrow \infty$ (see also
\cite{AS1}). Any improvements of this constant are very much
appreciated. Let us denote the conditional $\a$-probability of
$\cS$ given $\mathcal{PPT}$ as $\bbP_{\a}(\cS|\mathcal{PPT}, N,
D)=: \frac{V_{\a}(\cS)}{V_{\a}(\mathcal{PPT})}$ for fixed $\a>0$
and $N=D^n$. Below, we provide some numerical examples to see how
small of $\bbP_{\a}(\cS|\mathcal{PPT}, N, D)$ for (moderately)
large $N$. For instance, $\bbP_{1.1}(\cS|\mathcal{PPT}, 4096,
2)\leq 2.5\times 10^{-2721940}$ with $n=12$,
$\bbP_{1.1}(\cS|\mathcal{PPT}, 6561, 3 ) \leq 1.82\times
10^{-12248770}$ with $n=8$, and $\bbP_{0.9}(\cS|\mathcal{PPT},
6561, 3)\leq 3.74\times 10^{-735611}$ with $n=8$.

\section{Conclusion and Comments}

\vskip 2mm In the present paper, we compare several important
measures (Hilbert-Schmidt, Bures, $\alpha$-measure) on the set of
states $\cD$ on $\cH$. Roughly speaking, for any fixed $\a>0$ and
any measurable subset $\cK\subset \cD$, the $\a$-volume radii
ratio $\mathrm{VR}_{\a}(\cK, \cD)$ is bounded from below by
$\mathrm{VR}_{HS}(\cK, \cD)^{\max\{1,\a\}}$ and from above by
$\mathrm{VR}_{HS}(\cK, \cD)^{\min\{1,\a\}}$ (up to some universal
constants). We also compare the $\a$-volume radii ratio
$\mathrm{VR}_{\a}(\cK, \cD)$ with the Bures volume radii ratio
$\mathrm{VR}_{B}(\cK, \cD)$ and the $\beta$-volume radii ratio
$\mathrm{VR}_{\beta}(\cK, \cD)$ for fixed $\beta>0$. In
particular, Theorem \ref{Bures:a:compare} improves the estimates
on $\mathrm{VR}_{B}(\cK, \cD)$ in \cite{YeDeping2009} .

\vskip 3mm Employing these estimates to the set of separable
states $\cS$ and to $\mathcal{PPT}$ (the set of states with
positive partial transpose), we obtain both upper and lower bounds
for $\mathrm{VR}_{\a}({\cS, \cD})$ and $\mathrm{VR}_{\a}
(\mathcal{PPT}, \cD)$ for fixed $\a>0$. It is showed that
$\mathrm{VR}_{\a}(\mathcal{PPT}, \cD)$ is essentially a constant,
while $\mathrm{VR}_{\a}(\cS, \cD)$ is of order (at most) $N^{-k}$
for some $k>0$ (and so is $\mathrm{VR}_{\a}(\cS, \mathcal{PPT})$).
This means that a typical (PPT) quantum state is entangled;
namely, randomly choosing a (PPT) quantum state in $\cD$ on $\cH$
with (even moderately) large dimension, this state is entangled
with probability (very) close to $1$. This may be of importance in
analysis of quantum algorithms or quantum protocols that rely on
entanglement. This also shows that, for not-too-small dimensions,
the Peres-Horodecki PPT criterion is not a precise tool to
establish separability. Consequently, more precise necessary
and/or sufficient conditions for separability are in great demand.

\vskip 3mm In applications, one frequently considers $\cH$ as
$\cH=(\bbC^D)^{\otimes n}$, often with $D=2$ or $D=3$. The
estimates we obtain show that the $\alpha$-probability of $\cS$ in
$\cD$, or of $\cS$ in $\mathcal{PPT}$ is very small even for small
values of $n$. For example, the ratio of $2$-volumes
$\frac{V_2(\cS)}{V_2(\cD)}$ is less than $2.1\times 10^{-1595}$ if
$D=2$ and $n=8$, and similarly $\frac{V_2(\cS)}{V_2(\cD)}\leq
1.52\times 10^{-5301}$ if $D=3$ and $n=5$. Likewise, the ratio of
$(1.1)$-volumes $\frac{V_{1.1}(\cS)}{V_{1.1}(\mathcal{PPT})}\leq
2.5\times 10^{-2721940}$ if $D=2$ and $n=12$, and similarly,
$\frac{V_{1.1}(\cS)}{V_{1.1}(\mathcal{PPT})} \leq 1.82\times
10^{-12248770}$ if $D=3$ and $n=8$.

\vskip 3mm Concerning question (4), the situation is much less
clear. In fact, to answer this question, one needs to prove that,
for any bipartite system $\cH=\bbC ^D\otimes \bbC ^D$, there
exists a constant $C_0<1$, such that
$\mathrm{VR}_{\a}(\mathcal{PPT}, \cD)\leq C_0.$ Similar questions
can be asked by replacing $\a$-volume with the Hilbert-Schmidt
volume and the Bures volume.

\vskip 3mm \noindent {\bf Acknowledgement.} The author is grateful
to the reviewers and the editor for their valuable comments. The
author also would like to thank Dr.\ Szarek for reading the
manuscript carefully and for his valuable suggestions. This
research has been initiated with support from the NSF grants
DMS-0801275 and DMS-0652722, and completed while fully supported
by NSF-FRG grant: DMS-0652571.


\begin{thebibliography}{M-PK.2}
\small
\bibitem{K. Zyczkowski1998}
{ \.Zyczkowski K, Horodecki P, Sanpera A and Lewenstein M}  1998
 {\em Phys. Rev. A} {\bf 58} 883


\bibitem{RevModPhys.81.865}
{Horodecki R, Horodecki P, Horodecki M and Horodecki K}  2009
{\em Rev. Mod. Phys.} {\bf 81} 865


\bibitem{Nie1}
{  Nielsen  M A  and Chuang I L}  2000  {\em Quantum Computation
and Quantum Information} (Cambridge University Press)


\bibitem{EPR1}
{   Einstein A, Podolsky B and Rosen N}  1935  {\em Phys. Rev.}
{\bf 47} 777


\bibitem{Sch1}
{  Schr\"{o}dinger  E}  1935  {\em Die Naturwissenschaften} {\bf
23} 807


\bibitem{Gur}
{  Gurvits  L}  2004  {\em J. Comput. Syst. Sciences} {\bf 69} 448

\bibitem{Pe1}
{   Peres A}  1996  {\em Phys. Rev. Lett.} {\bf 77} 1413


\bibitem{Ho1}
{    Horodecki M, Horodecki P and Horodecki R}  1996  {\em Phys.
Lett. A} {\bf 223} 1


\bibitem{ST1}
{   St{\o}rmer E }  1963  {\em Acta Math.} {\bf 110} 233


\bibitem{Wor1}
{   Woronowicz S L}  1976  {\em Rep. Math. Phys.} {\bf 10} 165


\bibitem{Ho2}
{   Horodecki P}  1997  {\em Phys. Lett. A} {\bf 232} 333


\bibitem{AS1}
{   Aubrun G and Szarek S J}  2006  {\em Phys. Rev. A} {\bf 73}
022109



\bibitem{Sz1}
{  Szarek  S  J}  2005  {\em Phys. Rev. A}  {\bf 72} 032304


\bibitem{YeDeping2009}

{Ye D}  2009  {\em J. Math. Phys.} {\bf 50} 083502

\bibitem{Braunsein1996}

{Braunstein S L }  1996  {\em Phys. Lett. A}  {\bf 219} 169


\bibitem{Hall1}
{   Hall M J W }  1998  {\em Phys. Lett. A} {\bf 242} 123

\bibitem{Lloyd1988}

{ Lloyd S and  Pagels H} 1988 {\em Ann. Phys.} (N. Y.) {\bf 188}
186


\bibitem{Lubkin1978}
{Lubkin E}  1978  {\em J. Math. Phys.} {\bf 19} 1028


\bibitem{Page1993}
{Page D} 1993 {\em Phys. Rev. Lett.} {\bf 71} 1291



\bibitem{Sommers2004}
{   Sommers H J  and  \.Zyczkowski K} 2004 {\em J. Phys. A: Math.
Gen.} {\bf 37} 8457


\bibitem{Zycz2}
{\.Zyczkowski K and Sommers H J}  2001  {\em J. Phys. A} {\bf 34}
7111


\bibitem{Wer1}
{    Werner R F} 1989 {\em Phys. Rev. A} {\bf 40} 4277

\bibitem{Zycz1}
{   \.Zyczkowski K and Sommers H J} 2003 {\em J. Phys. A} {\bf 36}
10115

\bibitem{Deming1935}
{Deming W E  and Colcord C G} 1935  {\em Nature} (London) {\bf
135} 917



\bibitem{Ben1}
{   Bengtsson I  and \.Zyczkowski K} 2006 {\em Geometry of Quantum
States} (Cambridge University Press)


\bibitem{MML1}
{    Mehta M L} 1990 {\em Random matrices} 2nd edition (Academic
Press)

\bibitem{Ozawa2000}
{Ozawa M} 2000 {\em Phys. Lett. A} {\bf 268} 158

\bibitem{Soz1}
{   Sommers H J  and \.Zyczkowski K}  2003  {\em J. Phys. A} {\bf
36} 10083


\bibitem{Bur1}
{    Bures D J C}  1969  {\em Trans. Am. Math. Soc.} {\bf 135} 199

\bibitem{Uh0}
{   Uhlmann  A} 1976  {\em Rep. Math. Phys.} {\bf 9} 273

\bibitem{Barnum1996}

{ Barnum H, Caves C, Fuchs C, Jozsa R and Schumacher B}  1996 {\em
Phys. Rev. Lett.} {\bf 73} 2818

\bibitem{Petz1996}

{Petz D}  1996  {\em Linear Algebr. Appl.} {\bf 244} 81

\bibitem{Hip1}
{   Hiai F and  Petz D} 2000 {\em The semicircle Law, Free Random
Variables and Entropy} (American Mathematical Society,
Mathematical Surveys and Monographs, {\bf V. 77})


\bibitem{HLP}
{ Hardy G H,  Littlewood J E and  P\'{o}lya G} 1952 {\em
Inequalities} 2nd edition (Cambridge Universitu Press)

\end{thebibliography}
\end{document}